\documentclass[]{ICCFD6}

\usepackage{latexsym}
\usepackage{epsfig}
\usepackage{fancyhdr}
\usepackage{float}
\usepackage{amsmath}

\usepackage[C40,T1]{CJKutf8}
\newcommand{\chn}[1]{\bgroup\begin{CJK*}{UTF8}{}\CJKtilde\CJKfamily{gbsn}{#1}\end{CJK*}\egroup}

\newcommand{\qq}[1]{\lq{#1}\rq}
\newcommand{\pressure}{\ensuremath{\mathnormal{p}}}
\newcommand{\temperature}{\ensuremath{\mathnormal{T}}}
\newcommand{\zcoordinate}{\ensuremath{\mathnormal{z}}}
\newcommand{\zvelocity}{\ensuremath{\bar{\mathnormal{v}}_\mathnormal{z}}}

\begin{document}

\title{Molecular simulation of fluid dynamics on the nanoscale}

\author{Jadran Vrabec\footnote{Corresponding author: Prof.\ Dr.-Ing.\ habil.\ Jadran Vrabec $<$jadran.vrabec@upb.de$>$, phone +49 5251 602 421.}, Elmar Baumh\"ogger, Andreas Elsner, Martin Horsch,\\ Zheng Liu (\chn{刘峥}), Svetlana Miroshnichenko, Azer Nazdraji\'c \& Thorsten Windmann}

\addr{Thermodynamics and Energy Technology Laboratory (ThEt),\\
University of Paderborn, Warburger Str.\ 100, 33098 Paderborn, Germany}


\abstract{{\bf Abstract}: Molecular dynamics simulation is applied to Poiseuille flow of
   liquid methane in planar graphite channels, covering channel diameters between 3 and
   135 nm. On this length scale, a transition is found between the regime where local
   ordering induced by the wall dominates the entire system and larger channel diameters
   where the influence of boundary slip is still present, but of a more limited extent.
   The validity of Darcy's law for pressure-driven flow through porous media is
   not affected by the transition between these regimes.}

\maketitle

\pagestyle{empty}

%

\thispagestyle{empty}

\noindent
On the nanometer length scale, continuum approaches like the Navier-Stokes equation break
down, cf.\ Karnidiakis et al.\ (2005).
Therefore, the study of nanoscopic transport processes requires a molecular point
of view and preferably the application of molecular dynamics (MD) simulation.

In the past, MD could
be applied to small systems with a few thousand particles only,
due to the low capacity of computing equipment. Consequently,
a large gap between MD simulation results on the one hand and experimental
results as well as calculations based on continuum methods was present.
The constant increase in available computational
power is eliminating this barrier, and
the characteristic length of the systems accessible to MD simulation approaches
micrometers, see also Bernreuther and Vrabec (2005) for a discussion of efficient
massively parallel simulation algorithms and their implementation.

The present work deals with the flow
behavior of liquid methane, modeled by the truncated and shifted
Lennard-Jones (LJTS) potential, cf.\ Allen and Tildesley (1987),
confined between graphite walls. While the LJTS potential can also
be applied to the interaction between the solid wall and the fluid,
the carbon structure itself is modeled using a rescaled variant of
the Tersoff (1988) multi-body potential.
The unlike interaction parameters of the LJTS potential
acting between methane and carbon were determined
according to the Lorentz-Berthelot combination rule with the
Lennard-Jones parameters of Wang et al.\ (2000) for \qq{pure}
sp$^2$ configured carbon.

MD simulations of methane confined between graphite
walls with up to 4,800,000 interaction sites, i.e.\ carbon atoms and
methane molecules, were conducted while
the channel diameter was varied to include both the boundary-dominated regime
and the transition to the continuum regime.
A pressure gradient was induced by an external gravitation-like
acceleration acting on all methane molecules and a force in the
opposite direction acting on the carbon atoms of the graphite
wall. The flow was regulated using a proportional-integral controller
such that the wall velocity was zero while the fluid reached a
specified average velocity.

\begin{figure}[t!]
\centering
\includegraphics[width=7cm]{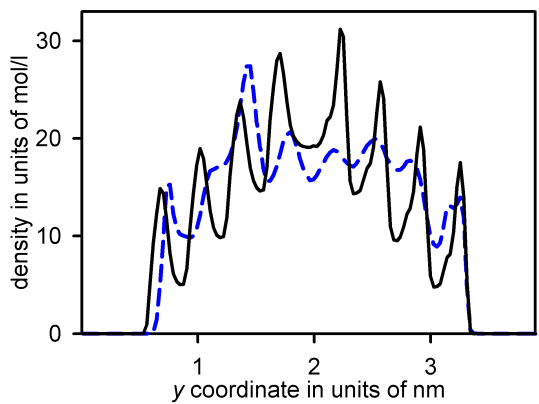}
\caption{
   Density profile for liquid methane at a temperature of 175 K in
   confined within a planar graphite channel with a diameter of 3 nm
   from MD simulation, averaged over the time intervals from 60 to
   120 ps (-- --) and from 420 to 480 ps (---) after simulation onset.
}
\label{0RF}
\end{figure}

\begin{figure}[h!]
\centering
\includegraphics[width=7cm]{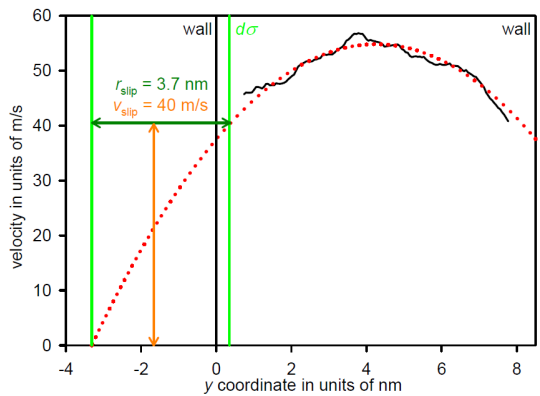}
\caption{
   MD simulation results $\mathrm{(\textnormal{---})}$ with a parabolic
   fit ($\cdot \cdot \cdot$) for the velocity profile during
   Poiseuille flow of liquid methane
   through a planar graphite channel with a diameter of 8.5 nm
   for an average flow velocity of 50 m/s at a
   density of 19 $\pm$ 1 mol/l in the central region
   of the channel and a temperature of 166.3 K.
}
\label{0RG}
\end{figure}

The fluid-solid interaction
induces a local ordering in the vicinity of the wall.
For channel diameters below 5 nm, cf.\ Fig.\ \ref{0RF},
this effect determines the structure of the entire system.
The resulting velocity profile is affected by the
local structure and therefore does not exhibit an exactly parabolic
shape. However, aggregated quantities such
as the slip velocity and the slip length, serving as boundary conditions for higher-order
CFD methods and in particular for Navier-Stokes solvers, can be determined
by extrapolating a parabolic fit as shown in Fig.\ \ref{0RG}.

\begin{figure}[b!]
\centering
\includegraphics[width=6.25cm]{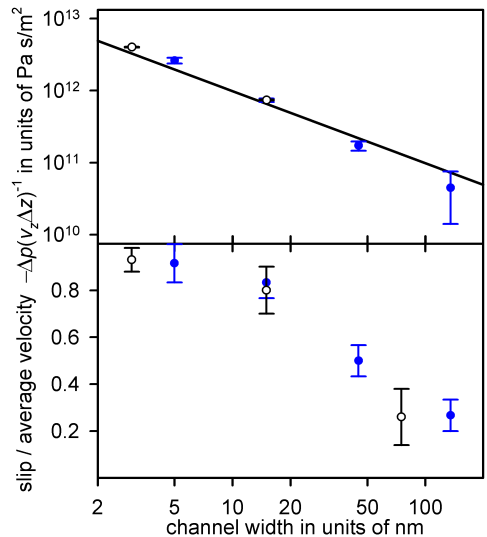}
\caption{Pressure drop $-\Delta\pressure$ in terms of $\zvelocity$ and
the channel length $\Delta\zcoordinate$ (top)
as well as slip velocity in terms of $\zvelocity$
(bottom), for Poiseuille flow of saturated liquid methane at
a temperature of $\temperature$ = 166.3 K
and average velocities $\zvelocity$ of 10 m/s (circles) and 30 m/s (bullets),
in dependence of the channel width; solid line: Darcy's law.
}
\label{0RDE}
\end{figure}

For channel diameters between 20 and 50 nm,
the boundary slip undergoes a qualitative transition, cf.\ Fig.\ \ref{0RDE}.
In an extremely narrow channel the
regular ordering of the fluid molecules due to the vicinity of the wall
entirely dominates not only the static structure, but also the fluid dynamics.
With respect to the characteristic
direction of the system, this highly ordered structure does
not support the extreme velocity gradient that would be implied by
the no-slip condition.

However, down to molecular length scales,
the pressure drop $-\Delta\pressure\slash\Delta\zcoordinate$
is approximately proportional to the average velocity $\zvelocity$
and inversely proportional to the cross-sectionional area of the channel,
cf.\ Fig.\ \ref{0RDE}, in agreement with Darcy's law.
Therefore, it can be concluded that
the qualitative transition between boundary-dominated laminar flow and 
laminar flow which is only influenced by boundary slip to a certain extent
is not reflected by a corresponding change for the
effective adhesive forces acting between the fluid and the solid.

The authors would like to thank M.\ Bernreuther (Stuttgart), M.\
Buchholz (Munich), J.\ Harting (Eindhoven), H.\ R.\ Hasse
(Kaisers\-lautern), S.\ Jakirli\'c (Darmstadt), and T.-H.\ Yen (Tainan)
for technical support as well as fruitful discussions and the German
Federal Ministry of Education and Research (BMBF) for funding the
project IMEMO. The presented research was carried out under the
auspices of the Boltzmann-Zuse Society of Computational Molecular
Engineering (BZS) and the simulations were performed on the Nehalem
cluster \textit{laki} at the High Performance Computing Center
Stuttgart (HLRS) under the grant MMSTP.

\end{document}